\documentclass{iopart}

\usepackage{amsfonts}
\usepackage{epsfig}

\renewcommand{\vec}[1]{\mathbf{#1}}
\newcommand{\hc}{\mathop{\rm h.\,c.}}

\newcommand{\sign}{\mathop{\mathrm{sgn}}}

\newcommand{\ep}{\epsilon}

\newcommand{\omph}{\omega_\mathrm{ph}}
\newcommand{\win}{\omega_\mathrm{in}}
\newcommand{\wout}{\omega_\mathrm{out}}
\newcommand{\photon}{\ell}
\newcommand{\nnvec}{\vec{d}}

\newcommand{\onlinecite}[1]{\cite{#1}}

\begin{document}

\title[Calculation of the Raman $G$ peak intensity in monolayer graphene]%
{Calculation of the Raman $G$ peak intensity in monolayer graphene:
role of Ward identities}
\author{D.~M.~Basko}
\address{Laboratoire de Physique et Mod\'elisation des Milieux Condens\'es,
Universit\'e Joseph Fourier and CNRS,
25 Rue des Martyrs, BP 166, 38042 Grenoble, France}
\ead{denis.basko@grenoble.cnrs.fr}


\begin{abstract}
The absolute integrated intensity of the single-phonon Raman peak at
$1580\:\mbox{cm}^{-1}$ is calculated for a clean graphene monolayer.
The resulting intensity is determined by the trigonal warping of the
electronic bands and the anisotropy of the electron-phonon coupling,
and is proportional to the second power of the excitation frequency.
The main contribution to the process comes from the intermediate
electron-hole states with typical energies of the order of the
excitation frequency, contrary to what has been reported earlier.
This occurs because of strong cancellations between different terms
of the perturbation theory, analogous to Ward identities in quantum
electrodynamics.
\end{abstract}

\maketitle

\section{Introduction}

In the past decades, Raman spectroscopy~\cite{Raman} techniques
were successfully applied to carbon compounds, such as
graphite~\cite{ThomsenReich2004} and carbon nanotubes~\cite{Dresselhaus2005}.
Upon the discovery of graphene~\cite{Novoselov2004}, Raman
spectroscopy has proven to be a powerful and non-destructive tool
to identify the number of layers, doping, disorder, strain, and
to characterize the phonons and electron-phonon coupling
\cite{Ferrari2006,Yan2007,Pisana2007,Das2008,Das2009,Mohiuddin2009,Heinz2009}.

The most robust feature in the Raman spectra of all conjugate
carbon compounds is the so-called $G$~peak, corresponding to the
in-plane bond-stretching optical vibration of $sp^2$-hybridized
carbon atoms. In graphene this vibration has the $E_{2g}$ symmetry,
it is doubly degenerate, and its frequency
$\omph\approx{1580}\:\mbox{cm}^{-1}$.
While the $G$~peak frequency is sensitive to external factors, such
as doping level~\cite{Yan2007,Pisana2007,Das2008}
or strain~\cite{Ni2008,Mohiuddin2009,Heinz2009},
its total frequency-integrated intensity $I_G$ is often assumed to
remain constant under the change of many external parameters,
depending only on the excitation frequency.
Thanks to this robustness, $I_G$ is often used as a reference to
which intensities of other peaks are compared
\cite{Elias2009,Heinz2008,Cancado2006}, since measurement of
absolute peak intensities represents a hard experimental task (the
first use of $I_G$ as a reference for other Raman peak intensities
in graphite probably dates back to 1970~\cite{Tuinstra}).

Given this popular role of $I_G$ as a reference, it is clear that
having at hand a theoretical expression of the absolute intensity
in terms of the basic parameters of the material (such as the
electronic dispersion, the electron-phonon coupling, \emph{etc.}),
would be useful. Nevertheless, to the best of the author's knowledge,
no such expression is available at present. $I_G$~was discussed in
a recent paper by the author~\cite{megapaper}, where the conclusion
was reached that even in the limit when the excitation frequency
$\win$ is small compared to the energy scale $t_0$ characterizing
the electronic dispersion (several eV), the intensity $I_G$ is
contributed by the whole conduction and valence bands, not just by
low-energy states in the vicinities of the Dirac points.
As will be shown below, this conclusion is wrong, since
Ref.~\onlinecite{megapaper} missed
strong cancellations occurring as a consequence of a Ward identity.
This affects the dependence of~$I_G$ on~$\win$.

In the present work $I_G$ is calculated for a clean graphene
monolayer suspended in vacuum. When $\win\ll{t}_0$, the intensity
is indeed determined only by electronic states with energies
$\sim\win$, not the whole conduction and valence bands (however,
these energies do not have to be close to $\win/2$).
In this limit $I_G$ can be expressed in terms of a few parameters,
characterizing these low-energy states. As correctly mentioned in
Ref.~\onlinecite{megapaper}, trigonal warping of the electronic
bands and the anisotropy of the electron-hole coupling turn out
to be crucial for the $G$~peak. In the limit $\omph\ll\win\ll{t}_0$
the frequency dependence is $I_G\propto\win^2$.
At higher frequencies, the calculation is performed using the
nearest-neighbor tight-binding model. The main feature found is a
strong enhancement of $I_G$ when the frequency matches the energy
of electron-hole separation at the $M$~point of the electronic first
Brillouin zone, where a van Hove singularity in the electronic
density of states occurs.

\section{Calculation}

\begin{figure}
\includegraphics[width=8cm]{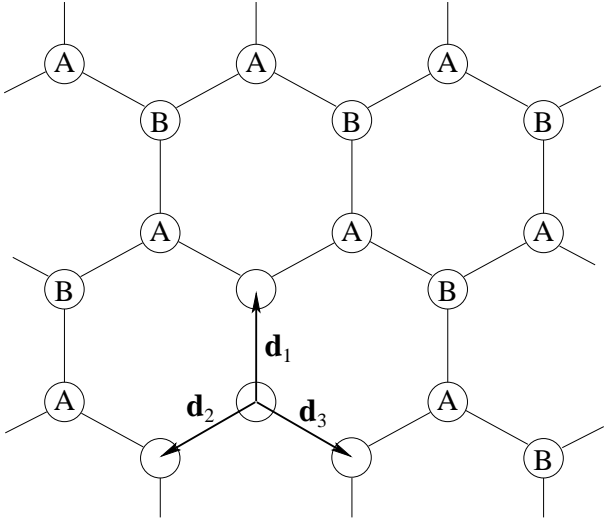}
\caption{\label{fig:lattice} Honeycomb lattice with the
two sublattices $A$ and $B$ and the three nearest-neighbor
bond vectors $\nnvec_1,\nnvec_2,\nnvec_3$.
}
\end{figure}

The Bloch form of the electronic wave function in the
tight-binding model is
\begin{equation}
\Psi_{\vec{k}}(\vec{r})=\rme^{{\rmi}\vec{k}\vec{r}}
\sum_{\alpha=A,B}\mathcal{U}_{\alpha\vec{k}}\sum_\vec{n}
\psi_\mathrm{a}(\vec{r}-\vec{r}_{\alpha,\vec{n}}),
\end{equation}
where $\psi_\mathrm{a}(\vec{r})$ is the atomic wave function
localized near the origin,
$\alpha=A,B$ labels the sublattices, and the two-dimensional
integer vector~$\vec{n}$ labels the unit cells, each
containing two atoms from the two sublattices. 
The wave vector~$\vec{k}$ runs over the first Brillouin
zone.
The Bloch amplitudes $\mathcal{U}_{\alpha\vec{k}}$ satisfy the
Shr\"odinger equation (we neglect the non-orthogonality of the
atomic orbitals)
\begin{equation}
\left[\ep-\hat{H}_\vec{k}^{(0)}\right]
\left(\begin{array}{c}\mathcal{U}_{A\vec{k}} \\ 
\mathcal{U}_{B\vec{k}}\end{array}\right)=0,
\end{equation}
where the Hamiltonian matrix is given by
\numparts
\begin{equation}\label{H0=}
\hat{H}_\vec{k}^{(0)}=
\left(\begin{array}{cc} 0 & {H}_{\vec{k}} \\ 
H_{\vec{k}}^* & 0 \end{array}\right),\quad
H_{\vec{k}}=-t_0\sum_{i=1,2,3}\rme^{{\rmi}\vec{k}\nnvec_i}
=H_{-\vec{k}}^*.
\end{equation}
Here $t_0\approx{3}\:\mbox{eV}$ is the nearest-neighbor
coupling matrix element, and $\nnvec_{1,2,3}$ are the
vectors connecting an $A$~atom to its three nearest neighbors,
as shown on Fig.~\ref{fig:lattice}. Their length is
$|\vec{d}_{1,2,3}|=a\approx{1}.42\:\mbox{\AA}$.
The energy eigenvalues are $\ep=\pm\ep_\vec{k}$, where
$\ep_\vec{k}\equiv|{H}_\vec{k}|$. Note that $H_\vec{k}$
is not a periodic function in the first Brillouin zone;
its values on the boundaries related by a reciprocal lattice
vector differ by a constant phase factor $\rme^{\pm{2}\pi{{\rmi}}/3}$.


Suppose that all atoms belonging to the sublattice~$\alpha$ are
displaced by $\vec{u}_\alpha$, and $\vec{u}_A=-\vec{u}_B$.
Such a uniform displacement corresponds to optical phonons with
the wave vector $\vec{q}=0$.
These are the phonons responsible for the $G$~peak, as fixed
by the momentum conservation (the wave vectors of both incident
and scattered photons are negligibly small). In the tight-binding
model the mechanism of the electron-phonon coupling is the change
of the nearest-neighbor electronic matrix element $t_0$
due to the change in the bond length~$a$,
$\partial{t}_0/\partial{a}\approx-6\:\mbox{eV}/\mbox{\AA}$.
Then, to linear order in the atomic displacements, the
electron-phonon interaction Hamiltonian is given by
\begin{eqnarray}\label{V0=}
\hat{V}_\vec{k}^{(0)}=
\frac{\vec{u}_A-\vec{u}_B}2\cdot\left(\begin{array}{cc} 0 & \vec{F}_{\vec{k}} \\ 
\vec{F}_{\vec{k}}^* & 0 \end{array}\right),\quad
\vec{F}_\vec{k}=\frac{2\rmi}{t_0a}\frac{\partial{t}_0}{\partial{a}}\,
\frac{\partial{H}_\vec{k}}{\partial\vec{k}}.
\end{eqnarray}

The electromagnetic field of the incident and scattered light is
described by the vector potential~$\vec{A}$. The Hamiltonian
describing interaction of the electrons with the field can be
obtained by the Peierls substitution $\vec{k}\to\vec{k}-(e/c)\vec{A}$
in the electronic Hamiltonian.
Since we are interested in a two-photon process, we should expand the
Hamiltonian to the second order in~$\vec{A}$:
\begin{eqnarray}
&&\hat{H}_\vec{k}^{(1)}=-\frac{e}c\,\vec{A}\cdot
\left(\begin{array}{cc} 0 & \vec{v}_{\vec{k}} \\ 
\vec{v}_{\vec{k}}^* & 0 \end{array}\right),\\
&&\hat{H}_\vec{k}^{(2)}=\frac{e^2}{c^2}\,A_iA_j
\left(\begin{array}{cc} 0 & w_{\vec{k}}^{ij} \\ 
(w_{\vec{k}}^{ij})^* & 0 \end{array}\right),\\
&&\hat{V}_\vec{k}^{(1)}=-\frac{e}c\,A_i(u_{Aj}-u_{Bj})
\left(\begin{array}{cc} 0 & g_{\vec{k}}^{ij} \\ 
(g_{\vec{k}}^{ij})^* & 0 \end{array}\right),\\
&&\hat{V}_\vec{k}^{(2)}=\frac{e^2}{c^2}\,A_iA_j(u_{Al}-u_{Bl})
\left(\begin{array}{cc} 0 & h_{\vec{k}}^{ijl} \\ 
(h_{\vec{k}}^{ijl})^* & 0 \end{array}\right),\\
&&\vec{v}_\vec{k}=\frac{\partial{H}_\vec{k}}{\partial\vec{k}},\quad
w_{\vec{k}}^{ij}=\frac{\partial^2{H}_\vec{k}}{\partial{k}_i\partial{k}_j},\quad
{g}_\vec{k}^{ij}=\frac{\partial{F}_\vec{k}^i}{\partial{k}_j},\quad
h_{\vec{k}}^{ijl}=\frac{\partial^2{F}^i_\vec{k}}{\partial{k}_j\partial{k}_l}.
\end{eqnarray}
\endnumparts
Here $i,j,l=x,y$ label the in-plane Cartesian components,
and the summation over repeated indices is assumed.
Note that for the description of resonant multiphoton
processes it is sufficient to keep $\hat{H}_\vec{k}^{(1)}$
term only, as it was done in~\cite{megapaper}, since the
contribution of other terms is (i)~off-resonant and
(ii)~smaller by a factor $\sim\win/t_0$. However, for the
one-phonon process studied here the contributions from all
terms turn out to be of the same order; in fact, they almost
cancel each other.

To calculate the Raman scattering probability, we follow the
general steps of Ref.~\onlinecite{megapaper}: pass to the
second quantization, calculate the scattering matrix
element, and substitute it into the Fermi Golden Rule.
The zero-temperature electronic Green's function is a
$2\times{2}$ matrix,
\begin{equation}\label{GF=}
\hat{G}_\vec{k}(\ep)=(\ep-\hat{H}_\vec{k}+\rmi{0}^+\sign\ep)^{-1}.
\end{equation}
Upon quantization of the phonon field the atomic displacements
corresponding to the optical phonons are expressed in terms of
the phonon creation and annihilation operators
$\hat{b}_{\vec{q},\mu}^\dagger,\hat{b}_{\vec{q},\mu}$ as
\begin{equation}
\hat{\vec{u}}_{A,\vec{n}}=-\hat{\vec{u}}_{B,\vec{n}}
=\sum_{\vec{q},\mu}\frac{\vec{e}^{(\mu)}}{\sqrt{2NM\omph}}
\left(\hat{b}_{\vec{q}\mu}\rme^{{\rmi}\vec{q}\vec{r}_{\vec{n}}}
+\hc\right).
\end{equation}
Here $\mu=x,y$ labels the two degenerate phonon modes,
$\vec{e}^{(\mu)}$ is the unit vector in the corresponding
direction, $M$~is the mass of the carbon atom, $N$~is
the total number of the carbon atoms in the crystal,
and ``h.c.'' stands for the hermitian conjugate.
A small finite wave vector $\vec{q}$ has to be introduced
in order to treat the in-plane momentum conservation
properly; only the $\vec{q}=0$ modes will contribute to
the final result. The transverse electromagnetic field is
quantized inside the large volume $L_xL_yL_z$ in the
Coulomb gauge:
\begin{equation}
\hat{\vec{A}}(\vec{r})=
\sum_{\vec{Q},\photon}\sqrt{\frac{2\pi{c}}{L_xL_yL_zQ}}
\left[\vec{e}^{(\vec{Q},\photon)}\hat{a}_{\vec{Q},\photon}\rme^{{\rmi}\vec{Q}\vec{r}}
+
\hc\right].
\end{equation}
Here $\vec{Q}$ is the three-dimensional wave vector,
$\ell=1,2$ labels the two transverse polarizations
along the two unit vectors
$\vec{e}^{(\vec{Q},\photon)}\perp\vec{Q}$.

\begin{figure}
\includegraphics[width=8cm]{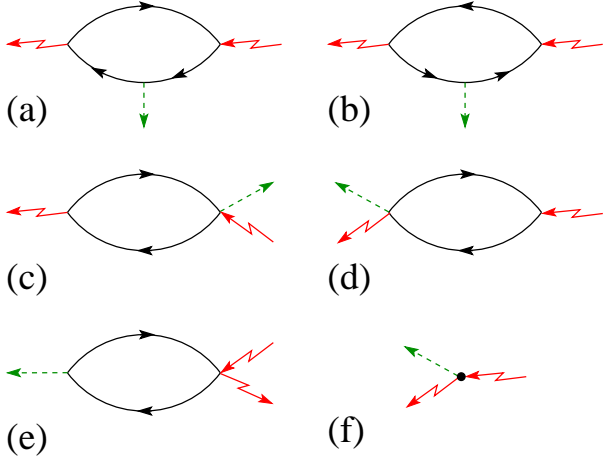}
\caption{\label{fig:diags} (Color on line.)
Diagrams corresponding to Eqs.~(\ref{diag1=})--(\ref{diag6=}).
Solid arrows correspond to the electronic Green's functions,
the dashed arrow represents the phonon emission vertex,
the zigzag arrows represent the photon absorption and
emission vertices.
}
\end{figure}

The probability for an incident photon with the polarization
$\vec{e}^\mathrm{in}$ and frequency $\win$ to be scattered into an
element of the solid angle $do_\mathrm{out}$ with the polarization
$\vec{e}^\mathrm{out}$, and with the frequency $\wout=\win-\omph$
fixed by the energy conservation, is given by
\begin{eqnarray}
&&\frac{dI_G}{do_\mathrm{out}}=\frac{\wout^2}{(2\pi)^2c^4}
\sum_\mu\left|2\mathcal{M}^{ijl}e^\mathrm{in}_i(e^\mathrm{out}_j)^*e^{(\mu)}_l\right|^2.
\end{eqnarray}
Here $\mathcal{M}^{ijl}$ is the transition matrix element
(the factor of 2 explicitly takes care of the two spin
projections of the electron). It is given by the sum of
the following contributions:
\begin{eqnarray}
&&\mathcal{M}^{ijl}=
\frac{2\pi{e}^2}{\sqrt{\win\wout}}\sqrt{\frac{L_xL_y}{2NM\omph}}
\int\frac{d^2\vec{k}}{(2\pi)^2}\frac{d\ep}{2\pi}\,\rme^{{\rmi}\ep{0}^+}
\nonumber\\ &&\quad\times
\Tr\{\mathcal{D}_1^{ijl}+\bar{\mathcal{D}}_1^{ijl}
+\mathcal{D}_2^{ijl}+\bar{\mathcal{D}}_2^{ijl}
+\tilde{\mathcal{D}}_2^{ijl}+\mathcal{D}_3^{ijl}\},\label{Mijl=}
\end{eqnarray}
\numparts
\begin{eqnarray}
&&\mathcal{D}_1^{ijl}=\hat{G}_\vec{k}(\ep)\hat{v}^i_\vec{k}\hat{G}_\vec{k}(\ep-\win)
\hat{F}^l_\vec{k}\hat{G}_\vec{k}(\ep-\wout)\hat{v}^j_\vec{k},\label{diag1=}\\
&&\bar{\mathcal{D}}_1^{ijl}=
\hat{G}_\vec{k}(\ep)\hat{v}^j_\vec{k}\hat{G}_\vec{k}(\ep+\wout)
\hat{F}^l_\vec{k}\hat{G}_\vec{k}(\ep+\win)\hat{v}^i_\vec{k},\label{diag2=}\\
&&\mathcal{D}_2^{ijl}=
\hat{G}_\vec{k}(\ep)\hat{g}^{li}_\vec{k}\hat{G}_\vec{k}(\ep-\wout)\hat{v}^j_\vec{k},\\
&&\bar{\mathcal{D}}_2^{ijl}=
\hat{G}_\vec{k}(\ep)\hat{v}^i_\vec{k}\hat{G}_\vec{k}(\ep-\win)\hat{g}^{lj}_\vec{k},\\
&&\tilde{\mathcal{D}}_2^{ijl}=
\hat{G}(\ep)\hat{w}^{ij}_\vec{k}\hat{G}_\vec{k}(\ep-\omph)\hat{F}^l_\vec{k},\\
&&\mathcal{D}_3^{ijl}=\hat{h}_\vec{k}^{lij}\hat{G}_\vec{k}(\ep).
\label{diag6=}
\end{eqnarray}\endnumparts
The factor $\rme^{{\rmi}\ep{0}^+}$ prescribes closing of the
$\ep$-integration contour in the upper half-plane. In fact,
it is important only for the last term, Eq.~(\ref{diag6=})
which corresponds to the sum over the filled valence band.
Upon integration over~$\ep$, Eq.~(\ref{Mijl=}) reduces
to the standard perturbation theory expression for the
Raman amplitude as the sum over intermediate states.
Each intermediate state contains an electron with the
wave vector~$\vec{k}$ in the conduction band and a hole
in the valence band with the wave vector $-\vec{k}$.
The $\vec{k}$~integration is performed over the first
Brillouin zone.
The six terms given by Eqs.~(\ref{diag1=})--(\ref{diag6=})
can be represented pictorially by the diagrams shown in
Fig.~\ref{fig:diags}.
In fact,
$\mathcal{D}_{1\vec{k}}=\bar{\mathcal{D}}_{1\vec{k}}$
because the Green's function and all the vertex matrices
satisfy
$\hat{G}_\vec{k}^T(\ep)=
\hat\sigma_x\hat{G}_\vec{k}(\ep)\hat\sigma_x$
and
$\hat{G}_\vec{k}(-\ep)=
-\hat\sigma_z\hat{G}_\vec{k}(\ep)\hat\sigma_z$
(the $T$ superscript denotes the matrix transpose,
and $\hat\sigma_{x,z}$ are the Pauli matrices).

If one integrates each of the terms in
Eqs.~(\ref{diag1=})--(\ref{diag6=}) separately,
the main contribution to the $\vec{k}$-integral
comes from the ``bulk'' of the first Brillouin zone,
rather than the vicinities of the Dirac points (as
it was pointed out in Ref.~\onlinecite{megapaper}).
Let us formally consider, however, the whole expression
for $\mathcal{M}^{ijl}$ at $\win=\wout=0$. Then the
sum of the terms (\ref{diag1=})--(\ref{diag6=}) is
a total derivative,
$\partial^2[\hat{F}_\vec{k}^l\hat{G}_\vec{k}(\ep)]/(\partial{k}_i\partial{k}_j)$.
Thus, the $\vec{k}$-integral vanishes, i.~e., at
$\win=\wout=0$ all terms contributing to the matrix
element, cancel. This cancellation is far more general
than the tight-binding model, used here. Indeed, consider
linear response of the electronic current~$\vec{j}$ to a
static homogeneous vector potential~$\vec{A}$. The
corresponding response function $\chi_{ij}$ must vanish,
since an observable (current) cannot respond to a pure
gauge%
\footnote{Strictly speaking, it is the limit
$\lim_{\vec{q}\to{0}}\lim_{\omega\to{0}}\chi_{ij}(\vec{q},\omega)$
which vanishes, while the calculation done in this work corresponds
to the opposite order of limits. The two limits commute for undoped
graphene, when the Fermi surface has zero length. At finite doping,
the contribution of the filled valence band states \emph{far below}
the Fermi energy still cancels out.
}.
This holds for any configuration of the atomic positions,
hence its derivative with respect to the displacements,
$\partial\chi_{ij}/\partial{u}_l$, must also vanish. As
follows from the Kubo formula, the matrix element
$\mathcal{M}^{ijl}$ at $\win=\wout=0$ is equal just to
this derivative, up to a constant factor. This fact is
analogous to the Ward identity in the quantum
electrodynamics. The cancellation of different terms
contributing to $\mathcal{M}^{ijl}$ was overlooked in
Ref.~\onlinecite{megapaper}, and it was wrongly concluded
that $\mathcal{M}^{ijl}$ is determined by the whole of
the first Brillouin zone. In fact, for $\win\ll{t}_0$
it is sufficient to focus on the vicinities of the two
Dirac points, as will be seen below.

The rest of the calculation is tedious, but straightforward.
Let us focus on the limit $\win\ll{t}_0$ first.
In the vicinity of the $K$ point we write
$\vec{k}=\vec{K}+\vec{p}$ and expand:
\numparts\begin{eqnarray}
&&H_\vec{k}=v(p_x-{\rmi}p_y)-\alpha_3(p_x+{\rmi}p_y)^2
+O(p^3),\label{Heff=}\\
&&\rmi\vec{F}_\vec{k}=F_0(\vec{e}^x-\rmi\vec{e}^y)
+F_1(p_x+{\rmi}p_y)(\vec{e}^x+\rmi\vec{e}^y)+O(p^2).
\label{Feff=}
\end{eqnarray}\endnumparts
The expansion around the $K'$~point can be obtained by
flipping $\vec{e}^x\to-\vec{e}^x$, $p_x\to-p_x$.
Note that going one order beyond the Dirac approximation
is necessary, since the leading contribution to
$\mathcal{M}^{ijl}$ vanishes.
(Indeed, the Dirac Hamiltonian has a continuous rotation
symmetry, so any third-rank tensor must vanish; the
continuous rotation symmetry is lifted by the trigonal
warping term.)
In the nearest-neighbor tight-binding model one has
\begin{equation}\label{TBrelations=}
v=\frac{3}2t_0a,\;\;\;
\alpha_3=\frac{3}8t_0a^2,\;\;\;
F_0=-3\frac{\partial{t}_0}{\partial{a}},\;\;\;
F_1=-\frac{3a}2\frac{\partial{t}_0}{\partial{a}}.
\end{equation}
Still, in Eqs.~(\ref{Heff=}), (\ref{Feff=}) we prefer
to keep all four parameters $v,\alpha_3,F_0,F_1$
independent, since going beyond the nearest-neighbor
tight-binding model would invalidate
Eq.~(\ref{TBrelations=}), while
the form of Eqs. (\ref{Heff=}), (\ref{Feff=}) is fixed
by the symmetry\footnote{
In fact, the $C_{6v}$ crystal symmetry allows a term
$\alpha_0(p_x^2+p_y^2)$ in the expansion of $H_\vec{k}$
(electron-hole asymmetry), which appears already in the
second-nearest-neighbor approximation. However, it does
not contribute to the matrix element calculated here
because of its full rotational symmetry.}.

It is instructive to write down explicitly the expression
for the matrix element after the integration over the
directions of~$\vec{p}$ (we change the integration variable
from~$p$ to $\ep_\vec{k}$ and denote by
$\mathcal{A}_C=L_xL_y/N=\sqrt{27}a^4/4$ the area per carbon atom):
\numparts\begin{eqnarray}
\mathcal{M}^{xxy}=\frac{{\rmi}e^2}{\sqrt{\win\wout}}
\sqrt{\frac{\mathcal{A}_C}{2M\omph}}
\int\limits_0^\infty{{d}\ep_\vec{k}}
\nonumber\\ \quad\quad\times
\left(\frac{C_\mathrm{in}\win^2}{4\ep^2_\vec{k}-\win^2}
+\frac{C_\mathrm{out}\wout^2}{4\ep^2_\vec{k}-\wout^2}
+\frac{C_\mathrm{ph}\omph^2}{4\ep^2_\vec{k}-\omph^2}\right)\nonumber\\
=-\frac{\pi e^2}{4}\sqrt{\frac{\mathcal{A}_C}{2M\omph}}
\frac{C_\mathrm{in}\win+C_\mathrm{out}\wout+C_\mathrm{ph}\omph}
{\sqrt{\win\wout}},\label{Mintegral=}\\
C_\mathrm{in}=2\,\frac{F_0\alpha_3}{v^2}\frac\win\omph
+\frac{F_1}{v}\left(\frac\win{2\omph}-\frac\win{2\wout}+1\right),\\
C_\mathrm{out}=-2\,\frac{F_0\alpha_3}{v^2}\frac\wout\omph
-\frac{F_1}{v}\left(\frac\wout{2\omph}+\frac\wout{2\win}-1\right),\\
C_\mathrm{ph}=-2\,\frac{F_0\alpha_3}{v^2}
+\frac{F_1}{v}\frac{\omph^2}{2\win\wout}.
\end{eqnarray}\endnumparts
The integral over $\ep_\vec{k}$ corresponds to the sum over the
intermediate states whose energies are $2\ep_\vec{k}$ (an electron
with the energy $-\ep_\vec{k}$ in the valence band is promoted to
the conduction band where its energy is $\ep_\vec{k}$). The poles
are bypassed by adding an infinitesimal imaginary part
$\ep_\vec{k}\to\ep_\vec{k}-i0^+$, as follows from Eq.~(\ref{GF=}).
%

\begin{figure}
\includegraphics[width=8cm]{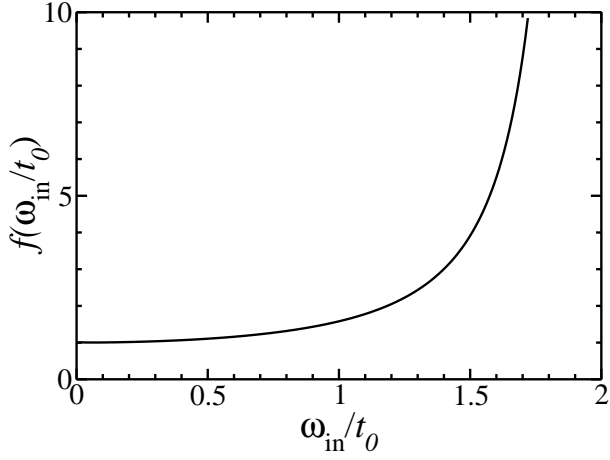}
\caption{\label{fig:plotf}
The dimensionless function $f(\win/t_0)$ appearing in Eq.~(\ref{IGfinal=}).
}
\end{figure}

At frequencies $\win$ not small compared to $t_0$ the
$\vec{k}$~integral in $\mathcal{M}^{xxy}$ should be evaluated
numerically. We focus in the limit $\omph\ll\win$. In this limit
the overall Raman efficiency, i.~e., the absolute probability
for an incident photon with the polarization $\vec{e}^\mathrm{in}$ and
frequency $\win$ to be scattered in the full solid angle $4\pi$
with any polarization, accompanied by the emission of a
$1580\:\mbox{cm}^{-1}$ optical phonon, is given by
\begin{equation}\label{IGfinal=}
I_G=\frac{2\pi\lambda_\Gamma}3\left(\frac{e^2}c\right)^2
\left(\frac{\win{a}}c\right)^2f(\win/t_0).
\end{equation}
The dimensionless electron-phonon coupling constant is
defined as
$\lambda_\Gamma=(\sqrt{27}/M\omph)%
(t_0^{-1}\partial{t}_0/\partial{a})^2$, which coincides
with the definition of Ref.~\onlinecite{megapaper} in
the tight-binding model.
The dimensionless function $f(\win/t_0)$ is plotted in
Fig.~\ref{fig:plotf}.
In the low-energy limit, $\win\ll{t}_0$, its value is $f(0)=1$,
as follows from Eq.~(\ref{Mintegral=}). When the frequency
matches the van Hove singularity at the $M$~point of the
first Brillouin zone, $\win\approx{2}t_0$, the intensity
diverges, $f(\win/t_0)\propto{1}/(\win/t_0-2)^2$.
This divergence is cut off at the scale
$|\win-2t_0|\sim\max\{\omph,\gamma\}$.
However, in the absence of a reliable information on electronic
relaxation processes in this region of the spectrum, we prefer
not to study this issue in detail, leaving the divergence as it
is.

\section{Discussion}

\begin{figure}
\includegraphics[width=\textwidth]{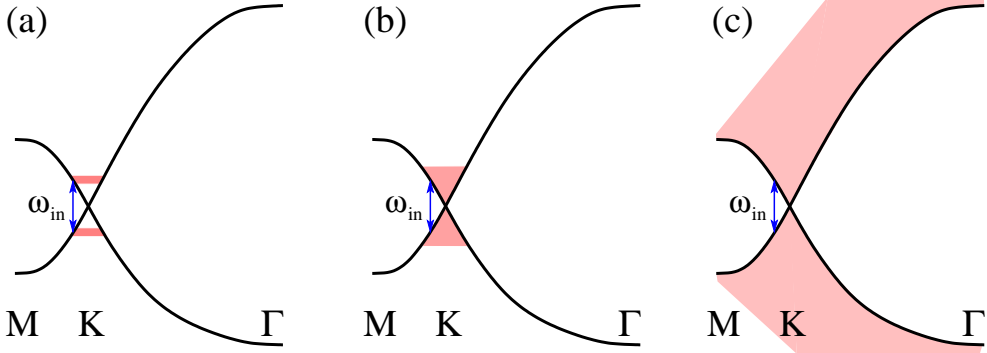}
\caption{\label{fig:band}
Schematic view of the $\pi$~electron band structure (solid
curves), with the states effectively contributing to a Raman
process (pink areas)for $\win\ll{t}_0$: (a)~states within
the narrow region $|\ep_\vec{k}-\win/2|\ll\win$;
(b)~a broad region of states with
$|\ep_\vec{k}-\win/2|\sim\win$, but still
$\ep_\vec{k}\ll{t}_0$;
(c)~the whole Brillouin zone, $\ep_\vec{k}\sim{t}_0$.
}
\end{figure}

First, let us see which electron-hole states contribute the most
to the $G$~peak intensity. In principle, for $\win\ll{t}_0$ one can
imagine three situations. They are illustrated Fig.~\ref{fig:band}
where the $\pi$~electron dispersion is shown schematically, together
with the contributing states.
In the situation~(a) these are the states whose energies are close
to the half of the excitation frequency, the difference
$\ep_\vec{k}-\win/2$ being small by some parameter.
This was shown to be the case for the $2D$~peak, the second-order
overtone of the $D$~peak at 2700~cm$^{-1}$, which is fully resonant
and $|\ep_\vec{k}-\win/2|\sim\gamma$, where $2\gamma\ll\win$ is the
electron inelastic scattering rate~\cite{shortraman, megapaper}.
This is also the case for the doubly-resonant $D$~peak at
1350 cm$^{-1}$, where $|\ep_\vec{k}-\win/2|\sim\omph\ll\win$
\cite{ThomsenReich,edgepaper}. In the situation~(b) the difference
$|\ep_\vec{k}-\win/2|\sim\win$ itself, and no small parameter enters.
Finally, in the situation~(c) the Raman process is contributed by the
whole Brillouin zone, $\ep_\vec{k}\sim{t}_0$.

A natural guess for the $G$~peak would be the option~(a): indeed,
there is a contribution from intermediate states that violate
energy conservation by the energy $\sim\omph$. However, in
Ref.~\cite{megapaper} it was found that the main contribution to
the integral in Eq.~(\ref{Mijl=}) with only two terms~(\ref{diag1=}),
(\ref{diag2=}) came from the states $\ep_\vec{k}\sim{t}_0$
[option~(c)].
In the present work we have seen that this contribution, in fact,
is canceled by other terms, missed in Ref.~\cite{megapaper},
due to the Ward identity.
To see that the correct picture for the $G$~peak in fact corresponds
to Fig.~\ref{fig:band}(c),
let us consider doped sample, i.~e., with the Fermi energy~$\ep_F$
detuned from the Dirac point. Then transitions involving states with
$\ep_\vec{k}<|\ep_F|$ are simply blocked by the Pauli principle, so
the lower limit of the integral in Eq.~(\ref{Mintegral=}) must be
set to $|\ep_F|$. Let us analyze the first term in
Eq.~(\ref{Mintegral=}) (the second term is almost equal to the first,
the third one is smaller by a factor $\sim\omph^2/\win^2$):
\begin{equation}
\int\limits_{|\ep_F|}^\infty
\frac{C_\mathrm{in}\win^2\,d\ep_\vec{k}}{4\ep^2_\vec{k}-\win^2}=
\frac{C_\mathrm{in}\win}{4}\left[i\pi\theta(\win-2|\ep_F|)
+\ln\left|\frac{2|\ep_F|+\win}{2|\ep_F|-\win}\right|\right].
\end{equation}
The value of the integral is determined entirely by the ratio
$|\ep_F|/\win$, and no other energy scales enter, whether
small ($\omph$) or large ($t_0$).
That is, the value of the integral in Eq.~(\ref{Mintegral=})
is determined not just by the vicinity of the pole, but by
the whole range of energies from $\ep_\vec{k}=0$
to $\ep_\vec{k}\sim\win$. 
Numerically, when $\ep_F$ is raised from zero to $\win/4$ (half of
the distance to the pole), $|\mathcal{M}^{xxy}|^2$ is changed by
about 12\%.

In other words, the uncertainty in the energy of the electron-hole
states, contributing to the process, is of the order of their energy
itself ($\sim\win$). By virtue of the energy-time uncertainty
principle, the duration of the process (the typical lifetime of the
virtual electron-hole pair) is $\sim{1}/\win$. The relevant length
scale is thus $v/\win\sim{3}\:\mbox{\AA}$ for
$v=10^8\:\mbox{cm/s}\approx{7}\:\mbox{eV}\cdot\mbox{\AA}$
and $\win=2\:\mbox{eV}$. The Raman process giving rise to the $G$~peak
is thus extremely local in space, involving just a few carbon atoms,
as it was noted earlier \cite{FerrariRobertson}. Its locality can
be also understood from the quasiclassical real-space picture of
Raman scattering in graphene \cite{edgepaper}.
This short length scale should be contrasted to much longer scales
responsible for Raman peaks produced according to
Fig.~\ref{fig:band}~(a), for example, the doubly-resonant $D$~peak
at 1350 cm$^{-1}$ \cite{Novotny,Gupta2009,Casiraghi2009,edgepaper}.

\begin{figure}
\includegraphics[width=8cm]{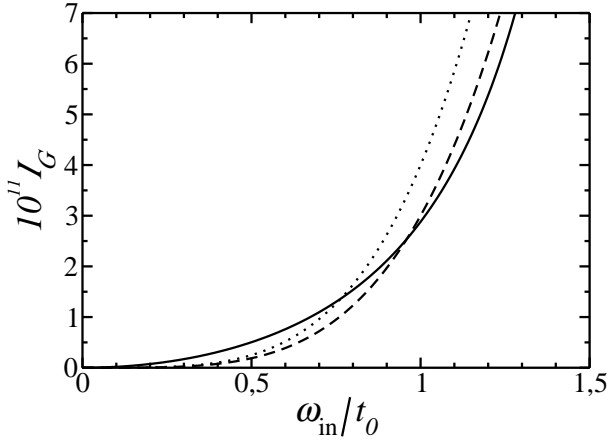}
\caption{\label{fig:omega4}
The intensity of the $G$~peak as a function of $\win/t_0$ from
Eq.~(\ref{IGfinal=}) for $\lambda_\Gamma=0.03$, $t_0a/c=0.0023$
(corresponding to $t_0\approx{3}.3\:\mbox{eV}$) (solid curve).
Dashed and dotted curves correspond to $3\cdot{10}^{-11}(\win/t_0)^4$
and $4\cdot{10}^{-11}(\win/t_0)^4$, respectively.
}
\end{figure}

Another consequence of the cancellation of the contribution from
$\ep_\vec{k}\sim{t}_0$ is the frequency dependence of $I_G$.
The standard textbook dependence $I\propto\omega^4$ (for
simplicity we assume $\win\approx\wout$) is obtained for
systems whose excited states have energies much higher
than~$\win$~\cite{RamanBook}. The $\omega^4$ dependence,
suggested in Ref.~\onlinecite{megapaper}, was essentially of
the same origin: it was assumed that the main contribution
to the process came from states in the bulk of the Brillouin
zone, and thus with high energies. As we have seen, this
conclusion is wrong, and the relevant states have energies
$\ep_\vec{k}\sim\win$. This remains true even at low
frequencies, as the electronic spectrum in graphene has no 
(or almost no) gap. As a result, at low frequencies the
dependence is $\propto\win^2$, as seen from Eq.~(\ref{IGfinal=}).
Nevertheless, a recent measurement of Raman peak intensities
performed on graphite nanocrystallites gives a $\win^4$
dependence \cite{Cancado2006,Cancado2007}.
For comparison, we plot $I_G$ from Eq.~(\ref{IGfinal=}) together
with two curves $\propto\win^4$ with two different coefficients
in Fig.~\ref{fig:omega4}.

Let us compare $I_G$ to the intensities of other peaks
which have been calculated theoretically and measured experimentally.
The $D$~peak at 1350 cm$^{-1}$ is due to emission of the
scalar $A_1$ phonons with wave vectors near the $K$~point. Due to
momentum conservation, the $D$~peak is activated by defects or edges.
For a defect-free graphene flake of the size~$L_a$ its
intensity is given by~\cite{edgepaper}
\begin{equation}
I_D=C\lambda_K\left(\frac{e^2}c\right)^2\frac{v^2}{c^2}
\frac{\win}{\omega_D^2}\frac{vL_a}{L_xL_y}
\ln\frac{\omega_D^2+(4\gamma)^2}{(4\gamma)^2}.
\end{equation}
Here  $\omega_D\approx{0}.17\:\mbox{eV}$ and $\lambda_K$ are the
frequency and the dimensionless electron-phonon coupling constant
for the corresponding phonons,
$v\approx{7}\:\mbox{eV}\cdot\mbox{\AA}$~is the electron velocity
(the slope of the Dirac cones),
$2\gamma$ is the electron inelastic scattering rate,
$L_xL_y$ is the area of the excitation laser spot, and
$C$~is a numerical coefficient which depends on the shape
of the flake and the character of the edges.
We take $\lambda_\Gamma=0.03$, as obtained from the doping
dependence of the $G$~peak frequency~\cite{Yan2007,Pisana2007},
and $\lambda_K/\lambda_\Gamma=3$ as extracted from the ratio of
intensities of the intensities of the
$2D$ peak at $2700\:\mbox{cm}^{-1}$ and the $2D'$~peak at
$3250\:\mbox{cm}^{-1}$~\cite{megapaper}.
Let us assume the inelastic scattering rate to be dominated by
electron-phonon scattering (which is a lower bound), then it can
be estimated as $\gamma=(\lambda_\Gamma+\lambda_K)\win/8$
\cite{shortraman}.
For a small flake, Eq.~(\ref{IGfinal=}) should be weighted
by the ratio of the area of the flake to the excitation spot area.
If we take an ideal hexagonal flake with armchair edges (which
definitely represents an upper bound for $I_D$),
then its area is $L_a^2\sqrt{27}/8$ and $C=4$.
For a round flake of the diameter~$L_a$ with atomically rough
edges the area is $\pi{L}_a^2/4$ and $C=\pi^2/18$, i.~e.,
$I_D$ is about 10 times smaller than for the ideal flake.
Taking $\win=2\:\mbox{eV}$ and using the low-frequency limit of
Eq.~(\ref{IGfinal=}), $f(\win/t_0)=1$, we obtain
$I_D/I_G\approx(300-3000\:\mbox{nm})/L_a$.
The experimentally found intensity is
$I_D/I_G=(560\:\mbox{nm}\cdot\mbox{eV}^4)/(L_a\win^4)$
\cite{Cancado2006,Cancado2007},
which at $\win=2\:\mbox{eV}$ gives an almost 10 times smaller
value than the calculated one for a flake with rough edges.

We can also consider $2D$~peak at $2700\:\mbox{cm}^{-1}$, whose
intensity was calculated to be \cite{shortraman,megapaper}
\begin{equation}\label{I2D=}
I_{2D}=\frac{\lambda_K^2}{24}\left(\frac{e^2}c\right)^2\frac{v^2}{c^2}
\frac{\win^2}{\gamma^2}.
\end{equation}
For the same values as above we obtain $I_{2D}/I_G\approx{150}$.
The experimental value for graphene samples on a substrate is
typically $I_{2D}/I_G\approx{5}$~\cite{Ferrari2006,Ni2008}.
For suspended samples the ratio can be
$I_{2D}/I_G\approx{9}$~\cite{Ni2009} and even
$I_{2D}/I_G\approx{17}$~\cite{Berciaud}.
Still, the experimental ratio is about 10 times smaller than
the calculated one.

Thus, the theory overestimates both $I_{2D}/I_G$ and $I_D/I_G$
by about an order of magnitude. It seems more reasonable to
assume that Eq.~(\ref{IGfinal=}) for~$I_G$ should be blamed for
this discrepancy, rather than expressions for $I_D$ and~$I_{2D}$.
Indeed, as discussed above, $I_D$ and $I_{2D}$ are determined by
just a few material parameters (electron-phonon coupling constants,
electronic velocity, etc.) which have been calculated by different
methods and measured in many independent experiments. Moreover,
corrections to Eq.~(\ref{I2D=}) due to the trigonal warping and
electron-hole asymmetry have been estimated in Ref.~\cite{megapaper}
and turned out to be small.
At the same time, the parameters determining $I_G$ (electronic
trigonal warping, electron-phonon coupling anisotropy) are not really
known, the validity of the tight-binding model at sufficiently high
energies is questionable, and even the calculated frequency dependence
of $I_G$ does not seem to agree with the experiment.

Still, what is the main source of such a strong discrepancy?
Is it possible that $\win=2\:\mbox{eV}$
is already sufficiently close to the singularity at $\win=2t_0$,
so that the dependence of $I_G$ on $\win$ is noticeably steeper than
$\omega^2$, and the value of $I_G$ is higher than that predicted
by the low-frequency asymptotics? On the one hand, the
nearest-neighbor tight-binding model with $t_0\approx{3}\:\mbox{eV}$
seems to describe the band structure reasonably well in the energy
range of interest, as is seen from its comparison with the results
obtained by angle-resolved photoemission spectroscopy~\cite{Bostwick2007}.
Then the energy of electron-hole separation at the $M$~point is
$2t_0\approx{6}\:\mbox{eV}$ and $\win/t_0=0.6$ is quite close to
the low-energy limit, as seen from Fig.~\ref{fig:plotf}.
On the other hand, a recent \textit{ab~initio} calculation of the
band structure of graphite gives the electron-hole separation at the
$M$~point about 4~eV~\cite{Gruneis2008}.
Also, the anisotropy of the electron-phonon coupling may be stronger
than that obtained from the nearest-neighbor tight-binding
model~\cite{Park2008}. In addition, the spectrum
of electron-hole excitations can be modified by excitonic effects:
the Coulomb attraction between the electron and the hole lowers
the energy of the excitation with respect to its non-interacting
value, so that the singularity in $I_D$ should be at a lower
frequency.
The author is aware of only one study of excitonic effects in the
optical response of monolayer graphene; within the Dirac model
it was shown that excitonic effects result in a weak feature at
the border of the electron-hole continuum~\cite{Mishchenko}.
Their importance for optical excitation of electrons near the
$M$~point of the first Brillouin zone remains to be studied.
Also, a direct measurement of the energy corresponding to the
singularity by an excitation in the ultraviolet frequency range
would shed light on this issue.

\section{Conclusions}

We have calculated the frequency-integrated intensity of the Raman
$G$~peak in monolayer graphene using the nearest-neighbor
tight-binding model for electrons. 
The resulting intensity is determined by the trigonal warping of the
electronic bands and the anisotropy of the electron-phonon coupling,
and is proportional to the second power of the excitation frequency.
Comparison to the intensities of other peaks, which have been
measured experimentally and calculated theoretically, suggests
that the present calculation underestimates $I_G$ by about an order
of magnitude.

\end{document}